 \documentclass{aa}  

\usepackage{graphicx}
\usepackage[varg]{txfonts}
\usepackage{natbib,twoopt}
\makeatother
\begin{document}

\title{A comparison of gyrochronological and isochronal age estimates for
transiting exoplanet host stars}

\titlerunning{Ages of transiting exoplanet host stars}
\author{P.~F.~L.~Maxted\inst{1}
 \and
A.~M.~Serenelli\inst{2}
 \and
J.~Southworth\inst{1}
 }

\institute{Astrophysics Group,  Keele University, Keele, Staffordshire
ST5~5BG, UK\\
 \email{p.maxted@keele.ac.uk}
\and
Instituto de Ciencias del Espacio (CSIC-IEEC), Facultad de Ciencias,
Campus UAB, 08193, Bellaterra, Spain
}

\date{Received ; accepted }

 
\abstract
{Tidal interactions between planets and their host stars are not well
understood, but may be an important factor in their formation, structure and
evolution. Previous studies suggest that these tidal interactions may be
responsible for discrepancies between the ages of exoplanet host stars estimated
using stellar models (isochronal ages) and age estimates based on the stars'
rotation periods (gyrochronological ages). Recent improvements in our
understanding of the rotational evolution of single stars and a substantial
increase in the number of exoplanet host stars with accurate rotation period
measurements make it worthwhile to revisit this question. }
{Our aim is to determine whether the gyrochronological ages for transiting
exoplanet host stars with accurate rotation period measurements are consistent
with their isochronal ages, and whether this is indicative of tidal
interaction between the planets and their host stars. }
{We have compiled a sample of 28 transiting exoplanet host stars with measured
rotation periods, including two stars (HAT-P-21 and WASP-5) for which the
rotation period based on the light curve modulation is reported here for the
first time. We use our recently-developed Bayesian Markov chain Monte Carlo 
method to determine the joint posterior distribution for the mass and age of
each star in the sample. We extend our Bayesian method to include a calculation
of the posterior distribution of the gyrochronological age that accounts for
the uncertainties in the mass and age, the strong correlation between these
values, and the uncertainties in the mass-rotation-age calibration.
}
{The gyrochronological age ($\tau_{\rm gyro}$) is significantly less than the
isochronal age for about half of the stars in our sample. Tidal interactions
between the star and planet are a reasonable explanation for this discrepancy
in some cases, but not all. The distribution of $\tau_{\rm gyro}$ values is
evenly spread from very young ages up to a maximum value of a few Gyr, i.e.,
there is no obvious ``pile-up'' of stars at low values  or very high of
$\tau_{\rm gyro}$  as might be expected if some evolutionary or selection
effect were biasing the age distribution of the stars in this sample..
There is no clear correlation between $\tau_{\rm gyro}$ and the strength of
the tidal force on the star due to the innermost planet. There is clear
evidence that the isochronal ages for some K-type stars are too large, and
this may also be the case for some G-type stars. This may be the result of
magnetic inhibition of convection. The densities of HAT-P-11 and WASP-84 are
too large to be reproduced by any stellar models within the observed
constraints on effective temperature and metallicity. These stars may have
strongly enhanced helium abundances. There is currently no satisfactory
explanation for the discrepancy between the young age for CoRoT-2 estimated
from either gyrochronology or its high lithium abundance, and the extremely old
age for its K-type stellar companion inferred from its very low X-ray flux. }
{There is now strong evidence that the gyrochronological ages of some transiting
exoplanet host stars are significantly less than their isochronal ages, but it
is not always clear that this is good evidence for tidal interactions between
the star and the planet. }
   \maketitle
%

\section{Introduction}

 Stars are born rotating rapidly and can then lose angular momentum if they
have a magnetised stellar wind. This observation leads to methods to estimate
the age of  single late-type stars from their  rotation period, a technique
known as gyrochronology \citep{2007ApJ...669.1167B}. It is unclear whether
exoplanet host stars can be considered as single stars in this context. If
significant orbital angular momentum is transferred by tides from the orbit of
a planet to the rotation of the host star (``tidal spin-up'') then the star
may rotate faster than a genuine single star of the same age, i.e., its
gyrochronological age ($\tau_{\rm gyro}$) will be an underestimate of its
true age. There is currently no quantitative global theory to calculate the
efficiency of tidal spin-up for stars with convective envelopes because the
dissipation of tidal energy by the excitation and damping of waves in
convective atmospheres is not fully understood  \citep{1975A&A....41..329Z,
1977A&A....57..383Z, 2009ApJ...696.2054G, 2014arXiv1411.3802D,
2014ARA&A..52..171O}.  Alternatively, the planet may disrupt the magentic
field geometry of the star and thereby reduce the efficiency of angular
momentum loss in the magnetised stellar wind, although this is only likely to
be a significant effect for F-type stars ($T_{\rm eff}>6000$\,K) with short
period planets \citep[$P<10$\,d,][]{2010A+A...512A..77L}.

 If a planet transits its host star then an analysis of the eclipse light
curve can yield an accurate estimate for the radius of the star relative to the
semi-major axis of the planet's orbit, $R_{\star}/a$, provided that the
eccentricity of the orbit is known. This estimate can be combined with
Kepler's laws to estimate the density of the host star
\citep{2003ApJ...585.1038S}. The density can be combined with estimates for
the effective temperature and metallicity of the star to infer a mass and age
for the star by comparison with stellar models. The age derived by comparing
the properties of a star to a grid of stellar models is known as the
isochronal age ($\tau_{\rm iso}$). It is not straightforward to calculate the
statistical error  on $\tau_{\rm iso}$ because it is strongly correlated with
the estimate of the stellar mass ($M_{\star}$) and the distribution can be
strongly non-Gaussian, e.g., for stars near the end of their main-sequence
lifetime the probability distribution for  $\tau_{\rm iso}$ given the
observed constraints (the posterior probability distribution) can be
bi-modal \citep{Maxted2015}. There are also systematic errors in stellar
models due to uncertainties in the input physics. For example, the efficiency
of energy transport by convection is usually parameterised by  the mixing
length parameter $\alpha_{\rm MLT}$ (the path length of a convective cell in
units of the pressure scale height) that is not known {\it a-priori} but that
can be estimated by finding the value of $\alpha_{\rm MLT}$ for which the
models match the observed properties of the Sun.  The initial helium
abundance for the Sun is estimated in a similar way. The radii of some M- and
K-type dwarf stars are not accurately predicted by these ``solar-calibrated''
stellar models \citep{1973A&A....26..437H, 1997AJ....114.1195P}. This ``radius
anomaly'' for low mass stars is an active research topic that is motivated by
a need to better understand exoplanet host stars and is driven by advances in
simulating convection in low-mass stars \citep{2008IAUS..252...75L} and
incorporating magnetic fields into stellar models \citep{2013ApJ...779..183F}.

 \citet{2009MNRAS.396.1789P} found suggestive empirical evidence for excess
rotation in stars with close-in giant planets (``hot Jupiters'') based on  a
sample of 28 stars.  For 3 of these stars the rotation period was
measured from the rotational modulation of the light curve by star spots. For
the other 25 stars the  rotation period was estimated from the projected
equatorial rotation velocity measured from rotation line broadening ($V_{\rm
rot}\sin i_{\star}$). For 12 of the stars, $V_{\rm rot}\sin i_{\star}$ was
below the instrumental resolution. It is difficult to make accurate estimates
of $V_{\rm rot}\sin i_{\star}$ in these cases because it requires the
deconvolution of the instrumental profile from the observed spectral line
profiles and the results are sensitive to details of the stellar atmosphere
models used to calculate the intrinsic stellar line profiles.
\citet{2014MNRAS.442.1844B} used a maximum likehood method to estimate the
ages of 68 transiting exoplanet host stars using 5 different grids of stellar
models and also applied four different methods to calculate the age using
gyrochronology. The rotation periods for 8 of the stars in that study were
based on rotational modulation of the light curve, the remainder were
estimated from $V_{\rm rot}\sin i_{\star}$. Estimates of the rotation period
based on $V_{\rm rot}\sin i_{\star}$ are generally much less precise than
those measured directly from the lightcurve and can be affected by systematic
errors when the line broadening due to rotation is small compared to other
poorly-understood line-broadening effects such as macro-turbulence. Additional
uncertainty is introduced by the unknown inclination of the star's rotation
axis, $i_{\star}$. \citeauthor{2014MNRAS.442.1844B} found a ``slight tendency
for isochrones to produce older age estimates'' but that the ``evidence for
any bias on a sample-wide level is inconclusive.'' 

 Planet host stars are an interesting test case for our understanding of the
tidal interactions because tidal spin-up will increase the star's magnetic
activity, which may increase the loss of rotational angular momentum through a
magnetised stellar wind.  The balance of these two effects may lead to a
quasi-stationary state in which the planet's orbit is stable in the long term
\citep{2015A&A...574A..39D}. However, if tidal interactions between stars and
planets are strong enough to efficiently destroy massive, short-period planets
then we might expect to only find hot Jupiters around young stars. To test
this hypothesis it would be useful in these cases to be able to estimate an
accurate isochronal age, because it is likely that the star will be spun-up
during the destruction of the planet and so its gyrochronological age will not
be an accurate estimate of its true age. Comparing the isochronal and
gyrochronological ages is not a straightforward problem because the
uncertainties on these two age estimates are strongly non-Gaussian and
$\tau_{\rm gyro}$ depends on the assumed stellar mass, so $\tau_{\rm iso}$ and
$\tau_{\rm gyro}$ are also correlated.

 In \citet{Maxted2015} we developed a Markov chain Monte Carlo (MCMC) method
that enables us to estimate the joint posterior probability distribution for
$\tau_{\rm iso}$ and $M_{\star}$ based on the observed density, effective
temperature (T$_{\rm eff}$)  and metallicity ([Fe/H]) of a star. In this
paper we select a sample of 28 transiting exoplanet host stars with accurately
measured rotation periods (Section 2).  For two of the stars in this sample
the rotation periods are new estimates based on the rotational modulation of
the light curves. We then describe how we have extended the method of
\citet{Maxted2015} to include a calculation of the posterior probability
distribution for $\tau_{\rm gyro}$  and present the results of applying this
technique to our sample of 28 planet host stars (Section 3). The implication
of these results for our understanding of tidal spin-up of exoplanet host
stars and the reliability of solar-calibrated stellar models is discussed in
Section 4 and our conclusions are given in Section 5.

\section{Sample definition and data selection}

 We have compiled a list of 28 transiting exoplanet host stars  for which the
rotation period has been measured directly from the rotational modulation of
the light curve due to star spots or, in the case of 55~Cnc, spectroscopic
variability in the Ca\,{\sc ii} H and K emission line fluxes. For all these
stars the existence of a transiting planet has been confirmed by multiple
radial velocity measurements based on high-resolution spectroscopy or
dynamical analysis of transit timing variations in multi-planet systems (e.g.,
Kepler-30). The observed properties of these stars are given in
Table~\ref{DataTable}.  For each star we list the rotation period, P$_{\rm
rot}$, the orbital period, P$_{\rm orb}$, the effective temperature, T$_{\rm
eff}$, the logarithmic surface iron abundance relative to the Sun, [Fe/H], the
mean stellar density, $\rho_{\star}$, the flux from the star at the top of the
Earth's atmosphere, $f_{\oplus}$, and the logarithm of the stellar luminosity,
$\log (L_{\star})$. 

 We have used only one or two sources of input data for this analysis rather
than attempting to reconcile or take the average of multiple studies for each
system.  We have excluded stars with masses that are too low to be covered by
our grid of stellar models $(<0.6 M_{\sun})$ or too high for
gyrochronology to be applicable $(>1.25 M_{\sun})$. We have also
excluded stars where the density estimate is based on a {\it Kepler}
long-cadance (LC) light curve only, because these density estimates appear to
be biased compared to the density estimated using asteroseismology
\citep{2013ApJ...767..127H}. 

\subsection{Luminosity measurements}
We can include the observed luminosity of the star ($L_{\star}$) as an
additional constraint in the analysis to derive the mass and age of the star.
For stars that have a trigonometrical parallax in \citet{2007A&A...474..653V}
with precision $\sigma_{\pi}/\pi \loa 0.1$ we calculate $L_{\star}$ using the
flux at the top of the Earth's atmosphere ($f_{\oplus}$) estimated by
integrating a synthetic stellar spectrum fit by least-squares to the observed
fluxes of the star. Optical photometry is obtained from the Naval Observatory
Merged Astrometric Dataset (NOMAD)
catalogue\footnote{\url{http://www.nofs.navy.mil/data/fchpix}}
\citep{2004AAS...205.4815Z}, the Tycho-2 catalogue \citep{2000A&A...355L..27H},
The Amateur Sky Survey \citep[TASS, ][]{1997AAS...190.3010D} and the Carlsberg
Meridian Catalog 14 \citep{2006yCat.1304....0C}. Near-infrared photometry was
obtained from the Two Micron All Sky Survey
(2MASS)\footnote{\url{http://www.ipac.caltech.edu/2mass}} and Deep Near
Infrared Survey of the Southern Sky
(DENIS)\footnote{\url{http://cdsweb.u-strasbg.fr/denis.html}} catalogues
\citep{2005yCat.2263....0T,2006AJ....131.1163S}. The synthetic stellar spectra
used for the numerical integration of the fluxes are from
\citet{1993KurCD..13.....K}. Reddening can be neglected for these nearby stars
given the accuracy of the measured fluxes and parallaxes. Standard errors are
estimated using a simple Monte Carlo method in which we generate  65,536 pairs
of $\pi$ and $f_{\oplus}$ values from Gaussian distributions and then find the
68.3\% confidence interval of the resulting $\log(L/L_{\sun})$ values.  The
2MASS photometry for 55~Cnc is not reliable so for this star we have used the
value of $f_{\oplus}$ from \citet{2013ApJ...771...40B}.

\begin{table*}
\caption{Observed properties of stars in our sample.}
\label{DataTable}
\centering
 \begin{tabular}{@{}lrrrrrrrr}
\hline
\hline
\noalign{\smallskip}
Star  &\multicolumn{1}{c}{P$_\mathrm{rot}$}& \multicolumn{1}{c}{P$_\mathrm{orb}$}
&\multicolumn{1}{c}{\mbox{T$_\mathrm{eff}$}}&\multicolumn{1}{c}{[Fe/H]}
&\multicolumn{1}{c}{$\rho_{\star}$} & \multicolumn{1}{c}{$f_{\oplus}$} &
\multicolumn{1}{c}{$\log(L_{\star})$} & Ref.\\
      &\multicolumn{1}{c}{[d]}& \multicolumn{1}{c}{[d]}
&\multicolumn{1}{c}{[K]}&
&\multicolumn{1}{c}{[$\rho_{\sun}$]} & \multicolumn{1}{c}{[pW\,m$^{-2}$]} &
\multicolumn{1}{c}{$[L_{\sun}]$} &\\
\noalign{\smallskip}
\hline
55 Cnc      & $39.00 \pm 9.00 $&   0.74 &$ 5234 \pm  30 $&$ +0.31 \pm 0.04 $&$ 1.084 ^{+0.040}_{-0.036} $&$  120.4 \pm  1.0 $&$ -0.244 \pm 0.009 $& 1, 2, 3, 4 \\
\noalign{\smallskip}
CoRoT-2     & $ 4.52 \pm 0.02 $&   1.74 &$ 5598 \pm  50 $&$ +0.04 \pm 0.05 $&$  1.362 \pm 0.064 $& & & 5, 6 \\
\noalign{\smallskip}
CoRoT-4     & $ 8.87 \pm 1.12 $&   9.20 &$ 6190 \pm  60 $&$ +0.05 \pm 0.07 $&$ 0.790 ^{+0.064}_{-0.161} $& & & 7, 8 \\
\noalign{\smallskip}
CoRoT-6     & $ 6.40 \pm 0.50 $&   8.89 &$ 6090 \pm  70 $&$ -0.20 \pm 0.10 $&$  0.929 \pm 0.064 $& & & 9, 8 \\
\noalign{\smallskip}
CoRoT-7     & $23.64 \pm 3.62 $&   0.85 &$ 5313 \pm  73 $&$ +0.03 \pm 0.07 $&$  1.671 \pm 0.073 $& & & 10, 11, 12 \\
\noalign{\smallskip}
CoRoT-13    & $ 13.00 ^{+5.00}_{-3.00} $&   4.04 &$ 5945 \pm  90 $&$ +0.01 \pm 0.07 $&$  0.526 \pm 0.072 $& & & 13, 8 \\
\noalign{\smallskip}
CoRoT-18    & $ 5.40 \pm 0.40 $&   1.90 &$ 5440 \pm 100 $&$ -0.10 \pm 0.10 $&$  1.090 \pm 0.160 $& & & 14, 6 \\
\noalign{\smallskip}
HAT-P-11    & $ 30.50 ^{+4.10}_{-3.20} $&   4.89 &$ 4780 \pm  50 $&$ +0.31 \pm 0.05 $&$  2.415 \pm 0.097 $&$    5.9 \pm  0.3 $&$ -0.590 \pm 0.035 $& 15, 8 \\
\noalign{\smallskip}
HAT-P-21    & $15.90 \pm 0.80 $&   4.12 &$ 5634 \pm  67 $&$ +0.04 \pm 0.08 $&$  0.700 \pm 0.150 $& & & 16, 12 \\
\noalign{\smallskip}
HATS-2      & $24.98 \pm 0.04 $&   1.35 &$ 5227 \pm  95 $&$ +0.15 \pm 0.05 $&$  1.220 \pm 0.060 $& & & 17, 17 \\
\noalign{\smallskip}
HD 189733   & $11.95 \pm 0.01 $&   2.22 &$ 5050 \pm  50 $&$ -0.03 \pm 0.05 $&$  1.980 \pm 0.170 $&$   27.5 \pm  1.4 $&$ -0.489 \pm 0.024 $& 18, 19 \\
\noalign{\smallskip}
HD 209458   & $10.65 \pm 0.75 $&   3.52 &$ 6117 \pm  50 $&$ +0.02 \pm 0.05 $&$  0.733 \pm 0.008 $&$   23.1 \pm  1.2 $&$  0.248 \pm 0.041 $& 20, 19 \\
\noalign{\smallskip}
Kepler-17   & $12.10 \pm 1.56 $&   1.49 &$ 5781 \pm  85 $&$ +0.26 \pm 0.10 $&$ 1.121 ^{+0.015}_{-0.034} $& & & 21, 6 \\
\noalign{\smallskip}
Kepler-30   & $16.00 \pm 0.40 $&  29.33 &$ 5498 \pm  54 $&$ +0.18 \pm 0.27 $&$  1.420 \pm 0.070 $& & & 22, 23, 22 \\
\noalign{\smallskip}
Kepler-63   & $ 5.40 \pm 0.01 $&   9.43 &$ 5576 \pm  50 $&$ +0.05 \pm 0.08 $&$ 1.345 ^{+0.089}_{-0.083} $& & & 24, 24 \\
\noalign{\smallskip}
Qatar-2     & $11.40 \pm 0.50 $&   1.34 &$ 4645 \pm  50 $&$ -0.02 \pm 0.08 $&$  1.591 \pm 0.016 $& & & 25, 25 \\
\noalign{\smallskip}
WASP-4      & $22.20 \pm 3.30 $&   1.34 &$ 5540 \pm  55 $&$ -0.03 \pm 0.09 $&$  1.230 \pm 0.022 $& & & 26, 6 \\
\noalign{\smallskip}
WASP-5      & $16.20 \pm 0.40 $&   1.63 &$ 5770 \pm  65 $&$ +0.09 \pm 0.09 $&$  0.801 \pm 0.080 $& & & 6 \\
\noalign{\smallskip}
WASP-10     & $11.91 \pm 0.05 $&   3.09 &$ 4675 \pm 100 $&$ +0.03 \pm 0.20 $&$ 2.359 ^{+0.053}_{-0.047} $& & & 27, 28, 29 \\
\noalign{\smallskip}
WASP-19     & $11.76 \pm 0.09 $&   0.79 &$ 5460 \pm  90 $&$ +0.14 \pm 0.11 $&$  0.885 \pm 0.006 $& & & 30, 31 \\
\noalign{\smallskip}
WASP-41     & $18.41 \pm 0.05 $&   3.05 &$ 5450 \pm 150 $&$ -0.08 \pm 0.09 $&$  1.270 \pm 0.140 $& & & 32, 32 \\
\noalign{\smallskip}
WASP-46     & $16.00 \pm 1.00 $&   1.43 &$ 5600 \pm 150 $&$ -0.37 \pm 0.13 $&$  1.240 \pm 0.100 $& & & 33, 33 \\
\noalign{\smallskip}
WASP-50     & $16.30 \pm 0.50 $&   1.96 &$ 5400 \pm 100 $&$ -0.12 \pm 0.08 $&$  1.376 \pm 0.032 $& & & 34, 35, 34 \\
\noalign{\smallskip}
WASP-69     & $23.07 \pm 0.16 $&   3.87 &$ 4700 \pm  50 $&$ +0.15 \pm 0.08 $&$  1.540 \pm 0.130 $& & & 36, 36 \\
\noalign{\smallskip}
WASP-77     & $15.40 \pm 0.40 $&   1.36 &$ 5500 \pm  80 $&$ +0.00 \pm 0.11 $&$ 1.157 ^{+0.016}_{-0.020} $& & & 37, 37 \\
\noalign{\smallskip}
WASP-84     & $14.36 \pm 0.35 $&   8.52 &$ 5300 \pm 100 $&$ +0.00 \pm 0.10 $&$  2.015 \pm 0.070 $& & & 36, 36 \\
\noalign{\smallskip}
WASP-85     & $14.64 \pm 1.47 $&   2.66 &$ 5685 \pm  65 $&$ +0.08 \pm 0.10 $&$  1.280 \pm 0.010 $& & & 38, 38 \\
\noalign{\smallskip}
WASP-89     & $20.20 \pm 0.40 $&   3.36 &$ 4960 \pm 100 $&$ +0.15 \pm 0.14 $&$ 1.357 ^{+0.069}_{-0.075} $& & & 39, 39 \\
\noalign{\smallskip}
\hline
\end{tabular}   
\tablebib{
 (1)~\citet{2000ApJ...531..415H};
 (2)~\citet{2014IAUS..293...52D};
 (3)~\citet{2005ApJS..159..141V};
 (4)~\citet{2013ApJ...771...40B};
 (5)~\citet{2009A&A...493..193L};
 (6)~\citet{2012MNRAS.426.1291S};
 (7)~\citet{2008A&A...488L..43A};
 (8)~\citet{2011MNRAS.417.2166S};
 (9)~\citet{2010A&A...512A..14F};
(10)~\citet{2010A&A...520A..53L};
(11)~\citet{2014A&A...569A..74B};
(12)~\citet{2012ApJ...757..161T};
(13)~\citet{2010A&A...522A.110C};
(14)~\citet{2011A&A...533A.130H};
(15)~\citet{2011ApJ...743...61S};
(16)~\citet{2011ApJ...742..116B};
(17)~\citet{2013A&A...558A..55M};
(18)~\citet{2008AJ....135...68H};
(19)~\citet{2010MNRAS.408.1689S};
(20)~\citet{2008ApJ...683L.179S};
(21)~\citet{2014ApJ...788....1B};
(22)~\citet{2012Natur.487..449S};
(23)~\citet{2012ApJ...750..114F};
(24)~\citet{2013ApJ...775...54S};
(25)~\citet{2014arXiv1406.6714M};
(26)~\citet{2011ApJ...733..127S};
(27)~\citet{2009MNRAS.398.1827S};
(28)~\citet{2009MNRAS.392.1585C};
(29)~\citet{2013MNRAS.430.3032B};
(30)~\citet{2013MNRAS.428.3671T};
(31)~\citet{2013MNRAS.436....2M};
(32)~\citet{2011PASP..123..547M};
(33)~\citet{2012MNRAS.422.1988A};
(34)~\citet{2011A&A...533A..88G};
(35)~\citet{2013MNRAS.431..966T};
(36)~\citet{2014MNRAS.445.1114A};
(37)~\citet{2013PASP..125...48M};
(38)~\citet{2014arXiv1412.7761B};
(39)~\citet{2014arXiv1410.6358H}.
}
\end{table*}

\subsection{Rotation periods}
\begin{figure}
\mbox{\includegraphics[width=0.45\textwidth]{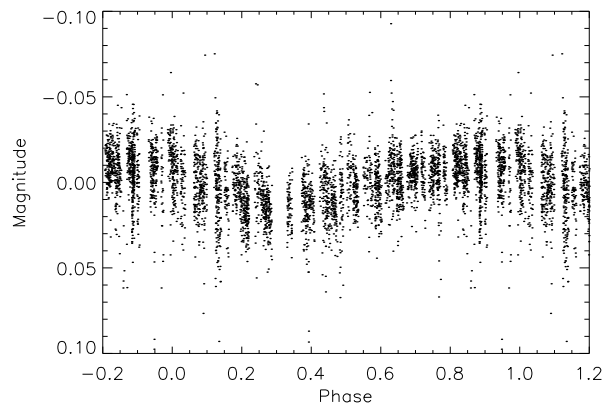}}
\caption{WASP photometry for HAT-P-21 plotted as a function of rotation phase
assuming $P_{\rm rot} = 15.9$\,d.
\label{ProtFig}}
\end{figure}

 The rotation periods for the stars in Table~\ref{DataTable} are taken from the
published sources noted except for HAT-P-21 and WASP-5, for which
we have measured the rotation periods using observations obtained by the WASP
project \citep{2006PASP..118.1407P} and the method described by
\citet{2011PASP..123..547M}. For HAT-P-21 we used 4185 observations obtained
between 2007~Jan~2 and 2007~May~15. These data show a very clear sinusoidal
signal with a period of 15.9\,d with an amplitude of 0.012 magnitudes. This
period agrees very well with the estimate $P_{\rm rot} = 15.7 \pm 2.2$\,d that
is expected based on the projected equatorial rotation velocity \citep[$V_{\rm
rot}\sin i_{\star} = 3.5 \pm 0.5 {\rm km\,s}^{-1}$, ][] {2011ApJ...742..116B} and the
radius of the star, assuming that the inclination of the star's rotation axis
is $i\approx 90^{\circ}$. The WASP data for HAT-P-21 are shown as a function
of rotation phase in Fig.~\ref{ProtFig}. The standard error on this value has
been estimated from the full-width at half-maximum of the peak in the
periodogram.

 For WASP-5 we analysed 6 sets of data obtained with 3 different cameras in 4
different observing seasons, as detailed in Table~\ref{WASP5Table}. In all 6
data sets we detect a periodic signal consistent with a rotation period
$P_{\rm rot} \approx 16$\,d if we allow for the possibility that the
distribution of star spots can produce a signal at the first harmonic of the
rotation period (i.e., at $P_{\rm rot}/2$). The value of $P_{\rm rot}$ in
Table~\ref{DataTable} is the mean of the observed values calculated with this
assumption and the error quoted in the standard error of the mean.

\citet{2013A&A...558A..55M} report a factor of 2 ambiguity in the rotational
period of HATS-2. We have used P$_{\rm rot}=24.98$\,d for our analysis since
this gives a value of $\tau_{\rm gyro}$ that is more consistent with the
value of $\tau_{\rm iso}$.

\begin{table}
 \caption{Periodogram analysis of the WASP light curves for \mbox{WASP-5}.
Observing dates are JD-2450000,  $N$ is the number of observations used in the
analysis, $A$ is the semi-amplitude of the sine wave fit by least-squares at
the period $P$ found in the periodogram with false-alarm probability FAP.
\label{WASP5Table}}
 \begin{tabular}{@{}lrrrrr}
\hline
  \multicolumn{1}{@{}l}{Camera} &
  \multicolumn{1}{l}{Dates} &
  \multicolumn{1}{l}{$N$} &
  \multicolumn{1}{l}{$P$ [d]} &
  \multicolumn{1}{l}{$A$ [mag]} &
  \multicolumn{1}{l}{FAP}\\
 \noalign{\smallskip}
\hline
\hline
227 & 3870\,--\,4054 & 4429 &   8.18 & 0.003 & 0.027 \\
228 & 3870\,--\,4020 & 3918 &  16.35 & 0.003 & 0.094 \\
228 & 4284\,--\,4433 & 4673 &   8.52 & 0.002 & 0.052 \\
225 & 5352\,--\,5527 & 7720 &  15.04 & 0.006 & $<0.001$ \\
225 & 5716\,--\,5897 & 4866 &  16.25 & 0.002 & 0.001 \\
228 & 5731\,--\,5885 & 3954 &  18.64 & 0.003 & 0.213 \\
 \noalign{\smallskip}
\hline
 \end{tabular}   
 \end{table}

\section{Analysis and Results}
\begin{figure*}
\mbox{\includegraphics[width=0.45\textwidth]{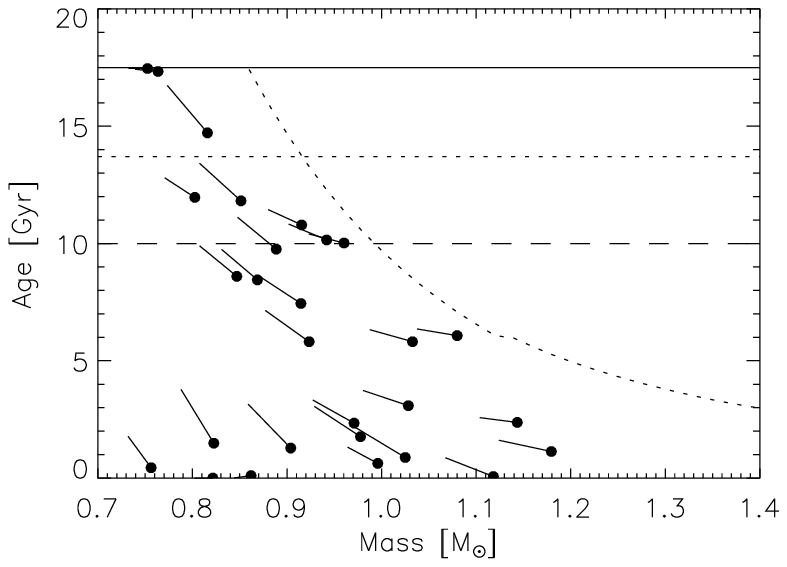}}
\mbox{\includegraphics[width=0.45\textwidth]{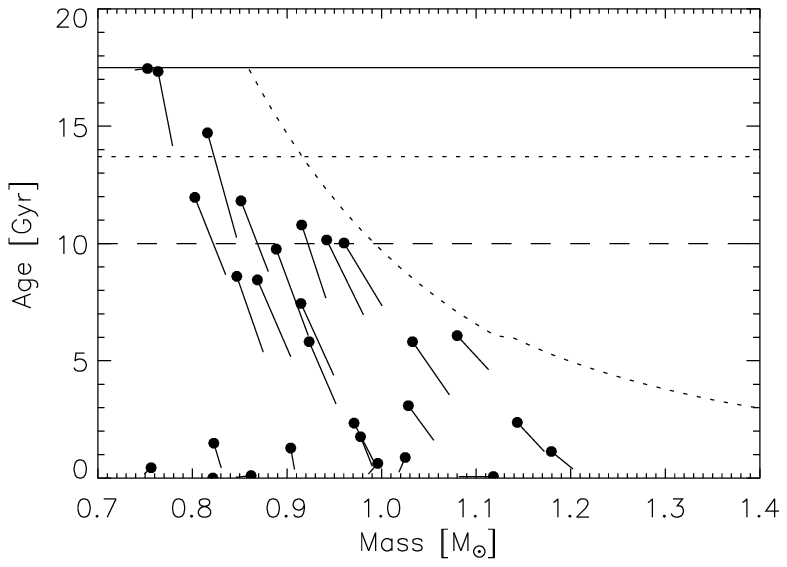}}
\caption{Change in the best-fitting masses and ages of transiting exoplanet
host stars due to a change in the assumed helium abundance or mixing length
parameter.  Dots show the best-fitting mass and age for the default values of
$Y$ and $\alpha_\mathrm{ MLT}$ and lines show the change in mass and age due
to an increase in helium abundance $\Delta Y = +0.02$ (left panel) or a change
in mixing length parameter $\Delta \alpha_{\rm MLT} = -0.2$ (right panel).
Horizontal lines indicate the age of the Galactic disc (dashed), the age of
the Universe (dotted) and the largest age in our grid of stellar models
(solid). The curved dotted line shows the terminal age main sequence (TAMS)
for stars with solar composition. \label{deltaFig}}
\end{figure*}

 We used version 1.1 of the program {\sc
bagemass}\footnote{\url{http://sourceforge.net/projects/bagemass}}
\citep{Maxted2015} to calculate the joint posterior distribution for the mass
and age of each star based on the observed values of T$_{\rm eff}$, [Fe/H],
the mean stellar density $\rho_{\star}$ and, if available, $L_{\star}$. The
stellar models used for our analysis were produced with the {\sc garstec}
stellar evolution code \citep{2008Ap&SS.316...99W}.  The initial
composition of the models is computed assuming a cosmic helium-to-metal
enrichment $\Delta Y/ \Delta Z = (Y_{\sun}-Y_\mathrm{BBN})/Z_{\sun}$, where
$Y_\mathrm{BBN}=0.2485$ is the primordial helium abundance due to big-bang
nucleosynthesis \citep{2010JCAP...04..029S}, $Z_{\sun} = 0.01826$ is the solar
metal abundance, and the initial solar helium abundance is $Y_{\sun}=0.26646$
so $\Delta Y/ \Delta Z = 0.984$. These models do not include rotation, but the
rotation rates of the stars in our sample are low (<2\% of breakup velocity in
all but two cases) and so it is not expected that rotation would play a direct
role in the structure of these stars. We do not rule out other, indirect
effects, e.g. low convective efficiency associated to large magnetic fields,
but these have to be modelled in an ad-hoc manner. This is discussed later in
this work, in relation to specific stars in the sample. The methods used to
calculate and interpolate the stellar model grid are described in
\citet{2013MNRAS.429.3645S} and \citet{Maxted2015}. We set lower limits of
80\,K on the standard error for T$_{\rm eff}$ and 0.07~dex for the standard
error on [Fe/H] \citep{2010MNRAS.405.1907B} and assume flat prior
distributions for the stellar mass and age. The results are given in
Table~\ref{ResultsTable}, where the maximum-likelihood (best-fit) values of
the stellar mass and age  are denoted $M_{\rm b}$ and $\tau_{\rm iso, b}$,
respectively. The mean and standard deviation of the posterior distributions
for the mass and age are listed under $\langle M_{\star} \rangle$ and $\langle
\tau_{\rm iso} \rangle$. Also listed in this table are our estimates of the
systematic errors in these values due to an assumed error of 0.2 for
$\alpha_{\rm MLT}$  and an assumed error of 0.02 for the initial helium
abundance, $Y$. The change in the estimated mass and age due to increasing $Y$
by its assumed error are given by the quantities $\sigma_{M, Y}$ and
$\sigma_{\tau, Y}$, respectively. Similarly,  $\sigma_{M,{\rm \alpha}}$ and
$\sigma_{\tau,{\rm \alpha}}$ quantify the change in the estimated mass and age
due to the error in $\alpha_{\rm MLT}$.  We show the best-fit values of
$M_{\star}$ and $\tau_{\rm iso}$ and the effects of changing $Y$ and
$\alpha_{\rm MLT}$ by their assumed uncertainties for all the stars in the
sample in Fig.~\ref{deltaFig}.

\subsection{Gyrochronological ages}
\begin{figure*}
\mbox{\includegraphics[width=0.45\textwidth]{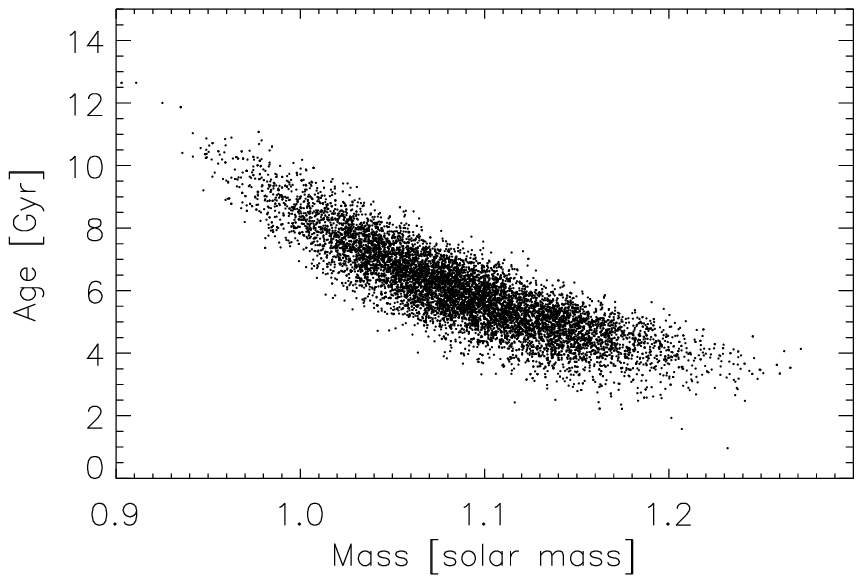}}
\mbox{\includegraphics[width=0.45\textwidth]{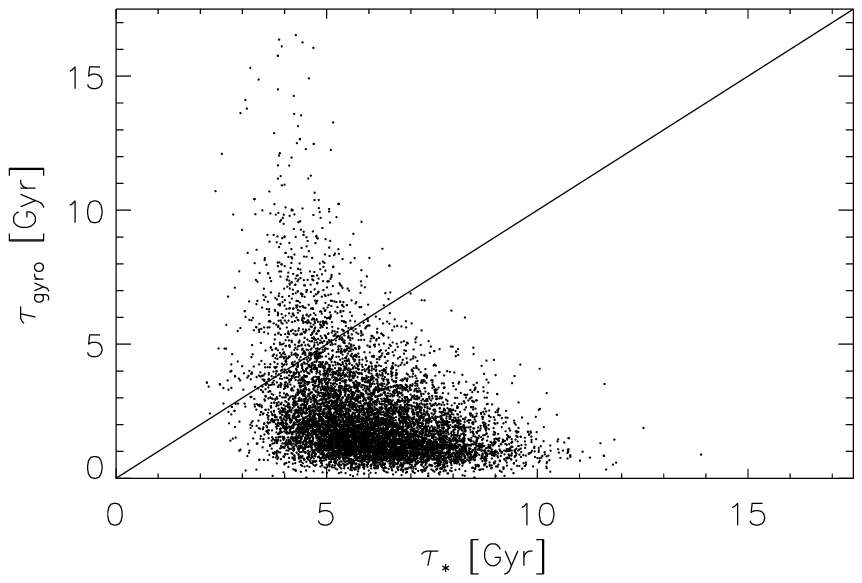}}
\caption{Left panel: Joint posterior distribution for the mass and age of CoRoT-13 estimated by our
isochrone fitting technique. Right panel: Joint posterior distribution  for
the age of CoRoT-13 estimated by our isochrone fitting technique and using
gyrochronology. For clarity, only 10\%  of the points from the Markov chain
are plotted. \label{CoRoT-13Fig}} \end{figure*}

 Our Bayesian MCMC method produces a large set of points in the mass-age
parameter space (``Markov chain'') that has the same distribution as the
posterior probability distribution for these parameters. We  used equation
(32) from \citet{2010ApJ...722..222B} to calculate $\tau_{\rm gyro}$ from the
rotation period and the mass of the star for every point in the Markov chain
for each star. The convective turn-over time scale  that encapsulates the
mass dependence of this mass-age-rotation relation was interpolated from
Table 1 of \citet{2010ApJ...721..675B}. This table suggests that stars with
masses \mbox{$M_{\star} \goa 1.25M_{\sun}$} do not have convective envelopes
so we have restricted our analysis to stars with
$P(M_{\star}<1.25M_{\sun})>0.9$ according to our MCMC results. We also checked
that none of the stars are close to the end of the main sequence, where rapid
changes in radius and internal structure are likely to make gyrochronological
ages unreliable. The calculation of $\tau_{\rm gyro}$ requires an estimate for
the value of the parameter $P_0$.  We account for the
uncertainty in this value by randomly generating  a value of $P_0$ uniformly
distributed in the range 0.12\,d to 3.4\,d for each point in the Markov chain.
In principle, there is some additional uncertainty in $\tau_{\rm gyro}$
due to the stars' surface differential rotation combined with the variation in
the latitude of the active regions that produce the star spot modulation
\citep{2014ApJ...780..159E}. We assume that this uncertainty affects the
calibration sample used by \citet{2010ApJ...722..222B} to the same extent that
it affects the stars in our sample so that this uncertainty is already
accounted for by randomly perturbing the parameter $P_0$. The uncertainties in
the observed values of $P_{\rm rot}$ are accounted for in a similar way, but
using Gaussian random distributions for the standard errors shown in
Table~\ref{DataTable}. Fig.~\ref{CoRoT-13Fig} shows the joint posterior
distributions for ($M_{\star}$, $\tau_{\rm iso}$) and ($\tau_{\rm iso}$,
$\tau_{\rm gyro}$) calculated from the Markov chain using this method for
CoRoT-13. 

\begin{figure} \mbox{\includegraphics[width=0.45\textwidth]{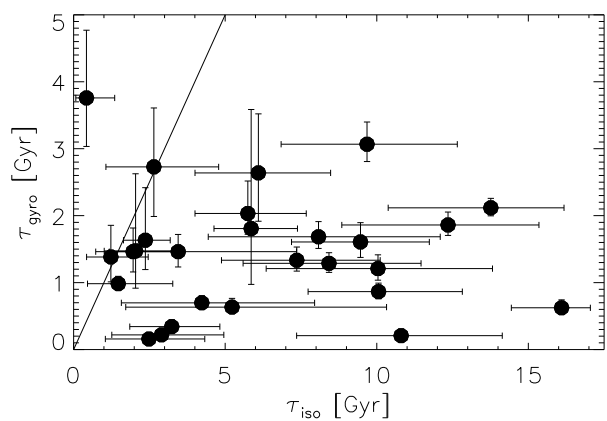}}
\caption{Comparison of gyrochronological ages ($\tau_{\rm gyro}$) to
isochronal ages ($\tau_{\rm iso}$) for planet host stars with measured
rotation periods. Points with error bars indicate the mean and standard
deviation of the posterior age distribution. The straight line is the relation
$\tau_{\rm gyro} = \tau_{\rm iso}$. \label{tgyroFig}}
\end{figure}

\begin{figure}
\mbox{\includegraphics[width=0.45\textwidth]{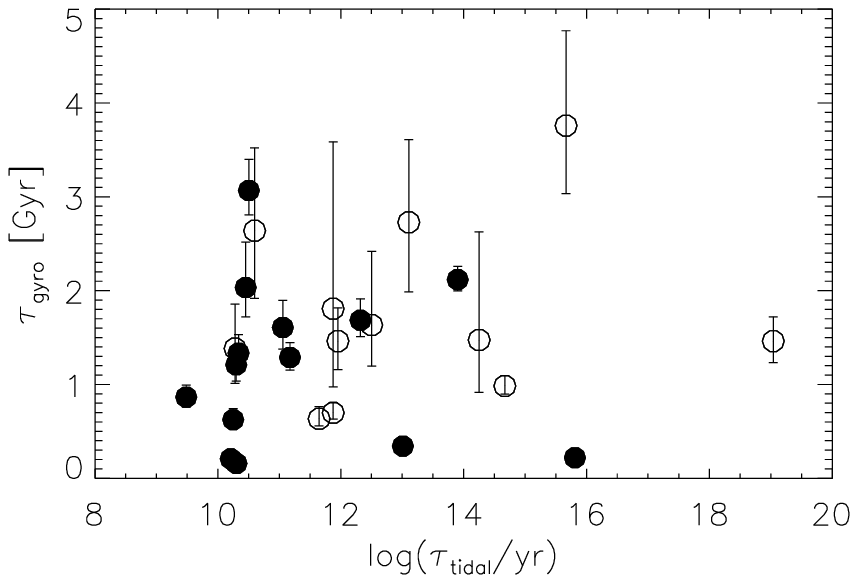}}
\caption{Gyrochronological ages as a function of $\log(\tau_{\rm tidal})$, a
very approximate estimate of the  time scale for tidal interaction between the
star and the innermost planetary companion. Note that the  time scale for
tidal interactions is uncertain by a few orders of magnitude. Systems with
significant differences between $\tau_{\rm iso}$ and $\tau_{\rm gyro}$
($p<0.05$) are plotted with filled symbols. 55~Cnc ($\tau_{\rm iso}\approx
9$\,Gyr, $\log(\tau_{\rm tidal}/{\rm y})\approx 13$) is not shown here.
\label{ttidalFig}}
\end{figure}

 For each star  we compare the values of $\tau_{\rm gyro}$ calculated using
the mass value at  every point in the Markov chain to the corresponding
$\tau_{\rm iso}$ values for the same point in the Markov chain. To quantify
the difference between $\tau_{\rm iso}$ and $\tau_{\rm gyro}$  we calculate
the fraction of points $p_{\tau}$ in the chain for which $\tau_{\rm gyro} >
\tau_{\rm iso}$, i.e., $p_{\tau} = P(\tau_{\rm gyro} > \tau_{\rm iso})$ is the
probability that the gyrochronogical age is greater than the isochronal age.
The results of this comparison are given in Table~\ref{tgyroTable} and are
shown in Fig.~\ref{tgyroFig}. Also given in Table~\ref{tgyroTable} is
$\tau_{\rm tidal}$, which is a very approximate estimate of the time scale for
tidal interactions between the star and the planet
\citep{2012ApJ...757...18A}. It must be emphasized that the actual time scale
for tidal spin-up in these systems is uncertain by a few order of magnitude
\citep{2014ARA&A..52..171O}. For multi-planet systems, $\tau_{\rm tidal}$
applies to the innermost planet. The values of $\tau_{\rm gyro}$ as a function
of $\tau_{\rm tidal}$ are shown in Fig.~\ref{ttidalFig}.

\begin{table*}
 \caption{Bayesian mass and age estimates for the host stars of transiting
extrasolar planets using {\sc garstec} stellar models assuming $\alpha_{\rm
MLT}=1.78$. Columns 2, 3 and 4 give the maximum-likelihood estimates of the
age, mass, and initial metallicity, respectively. Column 5 is the
chi-squared statistic of the fit for the parameter values in columns 2, 3, and
4.  Columns 6 and 7 give the mean and standard deviation of their posterior
distributions. Column 8 ($p_{\rm MS}$) is the probability that the star is
still on the main sequence. The systematic errors on the mass and age due to
uncertainties in the mixing length and helium abundance are given in columns 9
to 12.
\label{ResultsTable}}
 \begin{tabular}{@{}lrrrrrrrrrr}
\hline
\hline
Star &
  \multicolumn{1}{c}{$\tau_{\rm iso, b}$ [Gyr]} &
  \multicolumn{1}{c}{$M_{\rm b}$[$M_{\odot}$]} &
  \multicolumn{1}{c}{$\mathrm{[Fe/H]}_\mathrm{i, b}$} &
  \multicolumn{1}{c}{$\chi^2$}&
  \multicolumn{1}{c}{$\langle \tau_{\rm iso} \rangle$ [Gyr]}  &
  \multicolumn{1}{c}{$\langle M_{\star} \rangle$ [$M_{\odot}$]} &
  \multicolumn{1}{c}{$\sigma_{\tau, Y}$}  &
  \multicolumn{1}{c}{$\sigma_{\tau,\alpha}$} &
  \multicolumn{1}{c}{$\sigma_{M, Y}$}  &
  \multicolumn{1}{c}{$\sigma_{M,\alpha}$}  \\
\hline
 \noalign{\smallskip}
55 Cnc           &  10.9 &   0.91 &$ +0.378 $&  0.63 &$10.91 \pm  1.62 $&$  0.913 \pm 0.020 $&$  0.48 $&$  3.19 $&$ -0.033 $&$ -0.027 $ \\
CoRoT-2          &   1.8 &   0.97 &$ +0.057 $&  0.02 &$ 2.66 \pm  1.62 $&$  0.962 \pm 0.034 $&$  1.24 $&$  1.31 $&$ -0.047 $&$ -0.019 $ \\
CoRoT-4          &   1.2 &   1.17 &$ +0.076 $&  0.01 &$ 2.09 \pm  1.07 $&$  1.174 \pm 0.044 $&$  0.42 $&$  0.74 $&$ -0.046 $&$ -0.026 $ \\
CoRoT-6          &   3.2 &   1.02 &$ -0.160 $&  0.01 &$ 3.40 \pm  1.49 $&$  1.023 \pm 0.048 $&$  0.39 $&$  1.68 $&$ -0.037 $&$ -0.035 $ \\
CoRoT-7          &   1.3 &   0.90 &$ +0.046 $&  0.02 &$ 2.92 \pm  1.87 $&$  0.884 \pm 0.029 $&$  1.87 $&$  0.91 $&$ -0.045 $&$ -0.004 $ \\
CoRoT-13         &   6.1 &   1.08 &$ +0.083 $&  0.02 &$ 5.99 \pm  1.40 $&$  1.089 \pm 0.053 $&$  0.30 $&$  1.54 $&$ -0.046 $&$ -0.034 $ \\
CoRoT-18         &  11.9 &   0.85 &$ -0.011 $&  0.01 &$10.69 \pm  3.27 $&$  0.868 \pm 0.043 $&$  1.37 $&$  3.08 $&$ -0.046 $&$ -0.026 $ \\
HAT-P-11$^{\star}$&  0.0 &   0.82 &$ +0.238 $&  9.26 &$ 0.72 \pm  0.83 $&$  0.813 \pm 0.016 $&$  0.00 $&$ -0.00 $&$ -0.023 $&$  0.001 $ \\
HAT-P-21         &   9.9 &   0.96 &$ +0.121 $&  0.01 &$ 9.52 \pm  2.26 $&$  0.971 \pm 0.045 $&$  0.66 $&$  2.61 $&$ -0.045 $&$ -0.035 $ \\
HATS-2           &  10.0 &   0.89 &$ +0.214 $&  0.01 &$ 9.70 \pm  2.77 $&$  0.892 \pm 0.037 $&$  0.99 $&$  3.52 $&$ -0.037 $&$ -0.032 $ \\
HD 189733        &   1.6 &   0.82 &$ -0.014 $&  0.02 &$ 4.75 \pm  3.15 $&$  0.805 \pm 0.023 $&$  2.45 $&$  1.09 $&$ -0.036 $&$ -0.007 $ \\
HD 209458        &   2.4 &   1.14 &$ +0.065 $&  0.20 &$ 2.42 \pm  0.80 $&$  1.143 \pm 0.038 $&$  0.30 $&$  1.22 $&$ -0.043 $&$ -0.029 $ \\
Kepler-17$^{\star}$& 0.2 &   1.11 &$ +0.252 $&  0.04 &$ 1.48 \pm  1.07 $&$  1.075 \pm 0.034 $&$  0.53 $&$  0.10 $&$ -0.040 $&$  0.033 $ \\
Kepler-30        &   0.2 &   1.01 &$ +0.192 $&  0.01 &$ 4.38 \pm  3.24 $&$  0.919 \pm 0.065 $&$  1.23 $&$  0.16 $&$ -0.049 $&$  0.018 $ \\
Kepler-63        &   2.2 &   0.97 &$ +0.074 $&  0.00 &$ 3.16 \pm  1.88 $&$  0.958 \pm 0.037 $&$  1.08 $&$  1.58 $&$ -0.043 $&$ -0.023 $ \\
Qatar-2$^{\star}$&  17.5 &   0.75 &$ +0.149 $&  3.56 &$15.72 \pm  1.36 $&$  0.767 \pm 0.013 $&$  0.00 $&$  0.02 $&$ -0.018 $&$  0.013 $ \\
WASP-4           &   6.2 &   0.92 &$ +0.019 $&  0.04 &$ 6.27 \pm  2.34 $&$  0.919 \pm 0.045 $&$  0.95 $&$  2.85 $&$ -0.041 $&$ -0.031 $ \\
WASP-5           &   5.8 &   1.03 &$ +0.150 $&  0.00 &$ 5.84 \pm  1.86 $&$  1.035 \pm 0.048 $&$  0.62 $&$  2.25 $&$ -0.047 $&$ -0.038 $ \\
WASP-10          &   0.3 &   0.76 &$ +0.033 $&  0.03 &$ 6.00 \pm  4.12 $&$  0.709 \pm 0.031 $&$  2.56 $&$  0.18 $&$ -0.039 $&$  0.009 $ \\
WASP-19          &  10.0 &   0.94 &$ +0.222 $&  0.00 &$ 9.95 \pm  2.49 $&$  0.949 \pm 0.048 $&$  0.88 $&$  3.20 $&$ -0.044 $&$ -0.039 $ \\
WASP-41          &   8.4 &   0.87 &$ -0.022 $&  0.01 &$ 8.25 \pm  3.59 $&$  0.877 \pm 0.049 $&$  1.21 $&$  3.40 $&$ -0.039 $&$ -0.034 $ \\
WASP-46          &  12.1 &   0.80 &$ -0.294 $&  0.01 &$10.03 \pm  3.51 $&$  0.833 \pm 0.051 $&$  0.50 $&$  3.59 $&$ -0.030 $&$ -0.034 $ \\
WASP-50          &   8.6 &   0.85 &$ -0.054 $&  0.02 &$ 8.57 \pm  2.86 $&$  0.851 \pm 0.041 $&$  1.15 $&$  3.33 $&$ -0.038 $&$ -0.030 $ \\
WASP-69$^{\star}$&  17.5 &   0.76 &$ +0.251 $&  0.13 &$15.20 \pm  1.55 $&$  0.776 \pm 0.019 $&$ -0.04 $&$  2.94 $&$ -0.025 $&$ -0.015 $ \\
WASP-77          &   7.8 &   0.91 &$ +0.055 $&  0.04 &$ 7.57 \pm  2.53 $&$  0.918 \pm 0.049 $&$  0.46 $&$  3.10 $&$ -0.031 $&$ -0.034 $ \\
WASP-84$^{\star}$&   0.1 &   0.86 &$ -0.072 $&  0.73 &$ 1.89 \pm  1.61 $&$  0.826 \pm 0.025 $&$ -0.09 $&$  0.08 $&$ -0.010 $&$  0.013 $ \\
WASP-85          &   1.0 &   1.02 &$ +0.104 $&  0.02 &$ 2.10 \pm  1.38 $&$  0.994 \pm 0.036 $&$  1.12 $&$  0.71 $&$ -0.049 $&$  0.003 $ \\
WASP-89          &  14.9 &   0.81 &$ +0.236 $&  0.01 &$12.07 \pm  3.11 $&$  0.844 \pm 0.037 $&$  1.00 $&$  4.54 $&$ -0.033 $&$ -0.030 $ \\
 \noalign{\smallskip}
\hline
\end{tabular}   
$^{\star}$Best-fit is for age near the edge of the stellar model grid -- $\sigma_{\tau, Y}$ and $\sigma_{\tau,\alpha}$ may not be reliable.
\end{table*}     

\begin{table*}
 \caption{Stellar ages measured using our stellar models, $\tau_{\rm iso}$,
and using gyrochronology, $\tau_{\rm gyro}$, and the probability $p_{\tau} =
P(\tau_{\rm gyro} > \tau_{\rm iso})$. A very approximate estimate for the time
scale for tidal interaction between the star and the inner-most planet is
given under $\log(\tau_{\rm tidal})$. Notes include constraints on the age of
the star and any companion stars from the  X-ray luminosity ($\tau_{\rm X}$),
and age constraints from the   star's  surface lithium (Li) abundance
$\tau_{\rm Li}$
\label{tgyroTable}}
\begin{center}
 \begin{tabular}{@{}lrrrrl}
\hline
Star & \multicolumn{1}{l}{$\tau_{\rm iso}$}
&\multicolumn{1}{l}{$\tau_{\rm gyro}$}
& \multicolumn{1}{l}{$p_{\tau}$} 
&  $\log(\tau_{\rm tidal})$ 
& Notes  \\
&\multicolumn{1}{c}{[Gyr]}
&\multicolumn{1}{c}{[Gyr]}
& 
&\multicolumn{1}{c}{[yr]}  \\
\hline
\hline
 \noalign{\smallskip}
55 Cnc    &$ 10.91\pm 1.62$&$ 8.10 \pm  3.54$& 0.23& 12.8 & Companion star $\tau_{\rm X}> 5$\,Gyr. All age estimates consistent.\\
CoRoT-2   &$  2.66\pm 1.62$&$ 0.17 \pm  0.06$& 0.02& 10.3 & $\tau_{\rm Li}\approx 0.1$\,Gyr, $\tau_{\rm X}\approx 0.25$\,Gyr, but companion $\tau_{\rm X} >5$\,Gyr. \\
CoRoT-4   &$  2.10\pm 1.06$&$ 1.81 \pm  1.17$& 0.39& 14.3 & $\tau_{\rm iso}$ consistent with $\tau_{\rm gyro}$ but large relative error on both.\\
CoRoT-6   &$  3.40\pm 1.49$&$ 0.35 \pm  0.10$& 0.01& 13.0 & Fast rotator. \\
CoRoT-7   &$  2.92\pm 1.87$&$ 2.80 \pm  0.82$& 0.52& 13.1 & $\tau_{\rm iso}$ consistent with $\tau_{\rm gyro}$ but large relative error on both. \\
CoRoT-13  &$  5.99\pm 1.40$&$ 2.34 \pm  1.92$& 0.08& 11.9 & Tidal spin-up?\\
CoRoT-18  &$ 10.69\pm 3.28$&$ 0.22 \pm  0.08$& 0.00& 10.2 & Tidal spin-up?\\
HAT-P-11  &$  0.72\pm 0.83$&$ 3.89 \pm  0.89$& 0.98& 15.7 & Poor isochrone fit -- helium-rich?\\
HAT-P-21  &$  9.52\pm 2.26$&$ 1.64 \pm  0.29$& 0.00& 11.1 & $\tau_{\rm X} = 1$\,--\,2\,Gyr, companion  $\tau_{\rm X}>$ 5\,Gyr $\Rightarrow$ tidal spin-up. \\
HATS-2    &$  9.70\pm 2.77$&$ 3.10 \pm  0.30$& 0.01& 10.5 & Tidal spin-up?\\
HD 189733 &$  4.75\pm 3.15$&$ 0.71 \pm  0.09$& 0.07& 11.9 & $\tau_{\rm X} = 1$\,--\,2\,Gyr, companion  $\tau_{\rm X}>$ 5\,Gyr $\Rightarrow$ tidal spin-up. \\
HD 209458 &$  2.42\pm 0.79$&$ 1.83 \pm  0.85$& 0.29& 12.5 & $\tau_{\rm iso}$ consistent with $\tau_{\rm gyro}$ but large relative error on both.\\
Kepler-17 &$  1.48\pm 1.07$&$ 1.43 \pm  0.43$& 0.53& 10.3 & $\tau_{\rm iso}$ consistent with $\tau_{\rm gyro}$ but large relative error on both.\\
Kepler-30 &$  4.38\pm 3.24$&$ 1.47 \pm  0.24$& 0.22& 19.0 & $\tau_{\rm iso}$ consistent with $\tau_{\rm gyro}$ but large relative error on both.\\ 
Kepler-63 &$  3.16\pm 1.88$&$ 0.23 \pm  0.06$& 0.02& 15.8 & Fast rotator. \\
Qatar-2   &$ 15.72\pm 1.36$&$ 0.64 \pm  0.10$& 0.00& 10.2 & Inflated K-dwarf. \\
WASP-4    &$  6.27\pm 2.34$&$ 2.72 \pm  0.83$& 0.09& 10.6 & Tidal spin-up? \\
WASP-5    &$  5.84\pm 1.86$&$ 2.13 \pm  0.52$& 0.05& 10.5 & Tidal spin-up? \\
WASP-10   &$  6.00\pm 4.12$&$ 0.66 \pm  0.10$& 0.06& 11.6 & Tidal spin-up? \\
WASP-19   &$  9.95\pm 2.49$&$ 0.89 \pm  0.12$& 0.00&  9.5 & Tidal spin-up? \\
WASP-41   &$  8.25\pm 3.59$&$ 1.71 \pm  0.21$& 0.03& 12.3 & Magnetically active G-type star -- $\tau_{\rm iso}$ unreliable? \\
WASP-46   &$ 10.03\pm 3.51$&$ 1.23 \pm  0.20$& 0.01& 10.3 & Tidal spin-up? \\
WASP-50   &$  8.57\pm 2.86$&$ 1.30 \pm  0.15$& 0.00& 11.2 & Tidal spin-up? \\
WASP-69   &$ 15.20\pm 1.55$&$ 2.09 \pm  0.12$& 0.00& 13.9 & Inflated K-dwarf. \\
WASP-77   &$  7.57\pm 2.53$&$ 1.35 \pm  0.18$& 0.00& 10.3 & Companion $V_{\rm rot} \sin i_{\star}$ $\Rightarrow$ age $\approx$ 0.4\,Gyr. \\
          &$              $&$               $&     &      & Magnetically active G-type star -- $\tau_{\rm iso}$ unreliable? \\
WASP-84   &$  1.89\pm 1.61$&$ 0.99 \pm  0.10$& 0.36& 14.7 & Poor isochrone fit -- helium-rich?\\
WASP-85   &$  2.09\pm 1.37$&$ 1.50 \pm  0.33$& 0.41& 12.0 & $\tau_{\rm iso}$ consistent with $\tau_{\rm gyro}$ but large relative error on $\tau_{\rm iso}$. \\  
WASP-89   &$ 12.07\pm 3.11$&$ 1.88 \pm  0.18$& 0.00& 10.9 & Tidal spin-up? \\
 \noalign{\smallskip}
\hline
 \end{tabular}   
\end{center}
\end{table*}

\section{Discussion}

 It is clear that the gyrochronological age is significantly less than the
isochronal age for about half of the stars in this sample. This discrepancy is
apparent from Fig.~\ref{tgyroFig}, but the errors on these values are
correlated and non-Gaussian so to accurately quantify this discrepancy we need
to use the values of $p_{\tau}$ in Table~\ref{tgyroTable}.
 There is no obvious relation between the gyrochronological age of the stars
and the estimated time scale for tidal interactions between the star and the
planet. This may not be surprising given that $\tau_{\rm tidal}$  is uncertain
by a few orders of magnitude \citep{2014ARA&A..52..171O}. It is also apparent
that there is an even spread of $\tau_{\rm gyro}$ values from very young ages
up to a maximum value of a few Gyr, i.e., there is no obvious ``pile-up'' of
stars with very low or very high $\tau_{\rm gyro}$ values as might
be expected if some evolutionary or selection effect were biasing the age
distribution of the stars in this sample.

 The discrepancy between isochronal and gyrochronological ages for some
planet host stars has been noted for more limited samples of planet host
stars before, and has been cited as evidence for tidal spin-up of the stars by
their planetary companions, or for the disruption of the normal spin-down
process for these stars \citep{2009MNRAS.396.1789P, 2010A+A...512A..77L,
2014A&A...565L...1P}. However, there are other possibile explanations in some
cases. It may be that the isochronal ages for some stars are not accurate
because of missing physics in the stellar models. It is also possible for the
distribution of active regions on a star to produce a modulation of the light
curve with a period of half the rotation period. This is not a viable
explanation for the low values of $p_{\tau}$ in Table~\ref{tgyroTable} in
general, but may be an issue for a few stars with limited photometric data. We
have been careful to select stars for our sample for which the observational
data are robust, so the general appearance of distribution in
Fig.~\ref{tgyroFig} cannot be ascribed to observational errors, although there
may be issues with the observed data for a few stars. The limitations of the
observed data, the stellar models and the estimated value of $\tau_{\rm
tidal}$ affect different stars in this sample in different ways, so to
understand the implication of these results we need to look at the results on
a case-by-case basis.

\subsection{\object{HAT-P-11} and \object{WASP-84} -- helium
rich stars?}

 There is no satisfactory fit to the observed properties of HAT-P-11 at any
age or mass for our grid of standard stellar models, i.e. models using a
mixing length calibrated on the solar model ($\alpha_{\rm MLT}=1.78$) and
$\Delta Y/ \Delta Z = 0.984$, consistent with the primordial helium abundance
and  the assumed initial solar composition. In Table~\ref{HAT-P-11Table} we
summarize the published density estimates for HAT-P-11, including four studies
based on the very high quality {\it Kepler} short-cadence (SC) data for this
star. The {\it Kepler} data clearly show distortions to the light curve due to the
planet crossing dark spots on the face of the host star. Different
investigators have accounted for these spot-crossing events in different ways,
but the stellar densities derived are all consistent with the value we have
used in our analysis. There are also three independent analyses of the
spectrum of this star that have been used to derive the T$_{\rm eff}$ and
[Fe/H]. Again, all three values are consistent with each other, including the
value we have used for our analysis.

\begin{table*}
 \caption{Independent measurements of the mean density ($\rho_{\star}$),
effective temperature (T$_{\rm eff}$) and metallicity ([Fe/H]) for HAT-P-11
and their weighted mean values. \label{HAT-P-11Table}}
 \begin{tabular}{@{}rrrll}
  \multicolumn{1}{@{}l}{$\rho_{\star}/\rho_{\sun}$} &
  \multicolumn{1}{l}{T$_{\rm eff}$ [K]} &
  \multicolumn{1}{l}{[Fe/H]} &
 Source & Notes \\
\hline
 \noalign{\smallskip}
1.75$^{+0.92}_{-0.39}$ & $4780 \pm 50 $ &$  0.31 \pm 0.05 $ & \citet{2010ApJ...710.1724B}& Discovery paper \\ 
\noalign{\smallskip}
$2.42 \pm 0.10$  & & & \citet{2011MNRAS.417.2166S} & {\it Kepler} Q0\,--\,Q2\\ 
 \noalign{\smallskip}
$2.13^{+0.8}_{-0.5}$ & & &  \citet{2011ApJ...743...61S} & {\it Kepler} Q0\,--\,Q2\\
 \noalign{\smallskip}
$2.55 \pm 0.11 $& & &  \citet{2011ApJ...740...33D} & {\it Kepler} Q0\,--\,Q2 plus B-band and  J-band  \\
 & $4792 \pm 69$ & $0.33 \pm 0.07$ & \citet{2012ApJ...757..161T} & \\
 & $4624 \pm225$ & $0.26 \pm 0.08$ & \citet{2013A+A...558A.106M} & \\
$2.52 \pm  0.08$ & & &  \citet{2013A+A...560A.112M} & {\it Kepler} Q0\,--\,Q6\\
 \noalign{\smallskip}
\hline
 \end{tabular}   
 \end{table*}

 In principle, the {\it Kepler} short-cadence  photometry for HAT-P-11 can be used
to estimate the density of this star using asteroseismology.
\citet{2010ApJ...713L.164C} analysed the {\it Kepler} data for HAT-P-11 from the
commissioning period and the first month of regular observations and claimed
definite evidence for solar-like oscillations, yielding a preliminary estimate
of its mean density.  However, an analysis of the complete {\it Kepler} data set for
this star shows no convincing evidence for solar-like oscillations because
HAT-P-11 has a  much higher level of photometric noise than a typical star of
the same magnitude (Davies, priv. comm.).

 In general, the analysis of the light curve for a transiting extrasolar
planet makes the assumption that, apart from the effects of limb-darkening,
the mean surface brightness of a star is the same as the surface brightness in
the regions obscured by the planet. This would not be the case if, for
example, the planet transits a chord near  and approximately parallel to the
stellar equator of a star with dark spots near its poles. If the star spots
that are not occulted are equivalent to completely black spots that cover  a
fraction $f$ of the stellar disc, then from equation (19) of
\citet{2003ApJ...585.1038S}, we can estimate that the density will be
systematically too high by a factor $(1-f)^{-\frac{3}{4}}$. The density of
HAT-P-11 predicted by our stellar models based on the observed values of
T$_{\rm eff}$, $L_{\star}$ and [Fe/H]  assuming $\tau_{\rm iso} <10$\,Gyr is
$\rho_{\star}= 1.9\pm0.1\rho_{\sun}$. If this were to be explained by unocculted dark
spots this would require $f\approx 0.3$. Such a large value of $f$ can be
ruled out in the case of HAT-P-11 because of the unusual orbit of the planet
relative to the rotation axis of the star. The rotation and orbital axes are
almost perpendicular and the ratio of the orbital and rotation periods is
close to 1:6. This means the planet effectively scans the stellar disc along 6
different lines of longitude. Distortions to the light curve are seen due to
the planet crossing star spots with a typical size of about 5$^{\circ}$
located in two bands of latitude, similar to the pattern of spots seen on the
Sun \citep{2011ApJ...743...61S, 2014ApJ...788....1B}. These spots are too
small to explain the large value of $\rho_{\star}$ inferred from the transit
light curve of HAT-P-11.

 In principle, we could find a set of stellar models that match the properties
of HAT-P-11 by increasing the assumed mixing length parameter, but energy
transport by convection is expected to be less efficient in magnetically
active stars like HAT-P-11, not more efficient as a larger value of
$\alpha_{\rm MLT}$ would imply \citep{2013ApJ...779..183F}. A more plausible
explanation for the high density of HAT-P-11 is that this star has an
abnormally high helium abundance. We used {\sc garstec} to calculate  a grid
of stellar models identical to the grid of standard stellar models described
above, but with the helium abundance increased by an amount $\Delta Y =
+0.05$. With this ``very helium enhanced'' stellar model grid we find a good
fit to the observed properties of \mbox{HAT-P-11} ($\chi^2 = 0.34$) for a mass
$\langle M_{\star} \rangle = 0.74 \pm 0.02 M_{\sun}$ and an age $\langle
\tau_{\star} \rangle  = 1.7 \pm 0.7$\,Gyr. Clearly, there is some additional
systematic error  in these values because the actual helium abundance of
HAT-P-11 is unknown. It is very likely that other stars show variations in
their helium abundance around the simple linear relation between $Y$ and $Z$
that we have used. Another star in our sample with $\tau_{\rm iso,b}\approx 0$
and $\chi^2 \approx 0.5$ for $\nu=0$ degrees of freedom is WASP-84. It is
possible that this    is a very young star for which the observational errors
place the density slightly below the value predicted by our grid of standard
stellar models, but the results for HAT-P-11 also opens up the possibility that
this star has a more typical age and is helium-rich, particularly since it is
not associated with any star forming region and does not show any other signs
of extreme youth such as a high lithium abundance.

\subsection{\object{Qatar-2} and inflated K-dwarfs}
 It has long been known that some K-dwarfs  appear to be larger by 5\,per~cent
or more than the radius predicted by standard stellar models
\citep{1973A&A....26..437H, 1997AJ....114.1195P}. This  ``radius anomaly'' is
correlated with the rotation rate of the star, but also shows some dependence
on the mass and metallicity of the star \citep{2007ApJ...660..732L,
2013ApJ...776...87S}. The dependence on rotation is thought to be the result of
the increase in magnetic activity for rapidly rotating stars. Magnetic activity
can affect the structure of a star by producing a high coverage of starspots,
which changes the  boundary conditions at the surface of the star, or by
reducing the efficiency of energy transport by convection. Whatever the cause
of the radius anomaly in K-dwarfs, the existence of inflated K-dwarfs is one
likely explanation for  the appearance of stars in Table~\ref{ResultsTable}
that seem to be older than the Galactic disc
\citep[10\,Gyr,][]{2014A&A...566A..81C} or older than the Universe
\citep[13.75\,Gyr,][]{2013ApJS..208...19H}.

 One method that has been proposed to deal with the radius anomaly is to
simulate the magnetic inhibition of convection by reducing the mixing length
parameter  \citep{2007A&A...472L..17C}. This phenomenological approach has some
support from stellar models that incorporate magnetic fields in a
self-consistent way \citep{2013ApJ...779..183F}.  Qatar-2 is one K-dwarf in our
sample that has an isochronal age that is clearly older than the age of the
Galactic disc ($P(\tau_{\rm iso} < 10\,{\rm Gyr}) = 0.001$). We re-analysed the
data for Qatar-2 using grids of stellar models similar to those described above
but with various values of $\alpha_{\rm MLT}$ from 1.22 to 2.32. We find that
models with $\alpha_{\rm MLT} \loa 1.4$ can match the properties of this stars
for ages less than 10\,Gyr.

 The unknown value of $\alpha_{\rm MLT}$ is a source of systemic error that
potentially affects the mass and age estimates for all the stars that we have
studied. This includes the stars for which we have found a satisfatory fit to
the observations for some age within the range expected (0 to 10 Gyr), i.e., it
may be that we have assumed an inappropriate value of $\alpha_{\rm MLT}$ for
some stars but that this is hidden by deriving incorrect (but plausible) values
for the age and mass. 

\subsection{Stars with consistent isochronal and gyrochronological ages.}
From Table~\ref{tgyroTable} we see that there are nine stars for which there
is good agreement between $\tau_{\rm iso}$ and $\tau_{\rm gyro}$
($p_{\tau}>0.1$). These include the stars HAT-P-11 and WASP-84 discussed above
that may be helium-rich stars. As there is no good fit to the properties of
these stars within our model grid we do not discuss them further in this
section. 

 Of the seven remaining stars, HD~209458 and CoRoT-4 both have reasonably
precise gyrochronological and isochronal ages. For the other stars the
agreement between the two age estimates  is partly due to the large
uncertainty in one or both of these age estimates.  The time scale for tidal
spin-up of the star depends sensitively on the structure of the star and
whether there is any resonance between the orbit of the planet and internal
gravity wave modes in the star \citep{2014ARA&A..52..171O}, so we should not
expect that there is a precise value of $\log (\tau_{\rm tidal}/{\rm yr})$
that divides systems with and without tidal spin-up. Nevertheless, the good
agreeement between the isochronal and gyrochronological age for HD~209458 and
CoRoT-4 suggests that tidal spin-up for planet host stars may be inefficient
for systems with  $\log (\tau_{\rm tidal}/{\rm yr}) \ga 12.5$.

\subsection{Rapidly rotating stars}
 \cite{Maxted2015} compared the observed masses of stars in detached
eclipsing binaries to the masses predicted using {\sc bagemass} based on their
density, effective temperature and metallicity. They found that masses of some
stars with orbital periods less than about 6 days were under-predicted by
about 0.15\,$M_{\sun}$. The rotation periods of these stars are expected to be
equal to their orbital periods due to strong tidal interactions between the
stars. There were no stars in the sample of eclipsing binaries used by
\citet{Maxted2015} with orbital periods in the range 7\,--\,14 days, so it is
not known whether stars with rotation periods in this range are affected by
the same problem. The cause of this discrepancy is not known so the isochronal
ages for some stars with rotation periods less than about 6\,days will be
unreliable, and there is also the possibility that this problem affects stars
with rotation periods up to about 14\,days.

 There are four stars in our sample of planet host stars with rotation periods
less than about 6 days (CoRoT-2, CoRoT-6, CoRoT-18 and Kepler-63). All
four stars are clear examples of the gyrochronological age being significantly
lower than the isochronal age ($p_{\tau} < 0.02$).  Excluding these four stars
from the sample does not have a strong affect on our conclusions as there are
several other examples of stars with gyrochronological ages significantly  
lower than their isochronal age. 

 If we take the more cautious approach and exclude all stars in our sample
with rotation periods less than 14\,days and HAT-P-11 (discussed above), we
are left with a sample of 14 stars. Of these, about half have a
gyrochronological ages significantly  lower than their isochronal age
($p_{\tau} < 0.05$).

 CoRoT-6 and Kepler-63 are examples of stars where the tidal forces between
the star and the planet are very weak, e.g.,  weaker than the approximate
limit $\log (\tau_{\rm tidal}/{\rm yr}) \ga 12.5$ suggested in the previous
section, above which limit it is reasonable to assume that there is no tidal
spin-up for planet host stars. In these cases, we expect that the
gyrochronological age of the star should be reliable. If this is the case,
then it may be that these G-dwarfs stars are affected by magnetic inhibition
of convection in a similar way to K-dwarfs. WASP-41 is another  G-dwarf for
which the gyrochronological age  is significantly less than its isochronal age
($p_{\tau}=0.03$) despite having a very long tidal time scale ($\log
(\tau_{\rm tidal}/{\rm yr}) = 12.3 $). Although this is not a rapidly rotating
star ($P_{\rm rot} = 18.4$\,d) we mention it here because it is known to be a
magnetically active G8-type star based on the appearance of chromospheric
emission lines in its spectrum \citep{2011PASP..123..547M}. This is consistent
with the idea that some G-type stars have isochronal ages that are not
reliable (too large) as a result of magnetic activity that is related to the
rotation of the star but that is not directly  caused by rapid rotation.

\subsection{Comparison to other studies}

\citet{2014MNRAS.442.1844B} found a ``slight tendency for isochrones to
produce older age estimates'' but that the ``evidence for any bias on a
sample-wide level is inconclusive.''   All 8 stars in that sample with
directly measured rotation periods have been re-analysed here. For 7 of these
stars, our results are consistent with those of \citet{2014MNRAS.442.1844B}.
The exception is WASP-50, for which \citeauthor{2014MNRAS.442.1844B} uses a
value of T$_{\rm eff}$ based on photometric colour (Brown, priv. comm.),
rather than the lower and more accurate value that we have used here based on
an analysis of the spectrum from \citet{2011A&A...533A..88G}.  From a
comparison of the gyrochronological and isochronal ages,
\citeauthor{2014MNRAS.442.1844B} notes that of these 8 stars, 2 show ``an age
difference of a few Gyr''. We see very clear discrepancies between $\tau_{\rm
iso}$ and $\tau_{\rm gyro}$  for 7 of these 8 stars, i.e., $p_{\tau}<0.1$,
often much lower. We have restricted our analysis to stars with directly
measured $P_{\rm rot}$ values and increased the number of such stars studied
to 28 so we very clearly see that $\tau_{\rm iso}$ and $\tau_{\rm gyro}$
disagree for the majority of stars common to both studies. A key difference
between our study and the study by \citeauthor{2014MNRAS.442.1844B} is the way
that the uncertainties on the ages have been calculated.
\citeauthor{2014MNRAS.442.1844B} calculated the range of isochronal ages
corresponding to the ``1-$\sigma$'' error ellipse on T$_{\rm eff}$ and
$\rho_{\star}$ (the error on  [Fe/H] was neglected) and then used this range
as the standard deviation of a normal distribution to represent the posterior
probability distribution for $\tau_{\rm iso}$. The correlation between
$\tau_{\rm iso}$ and $\tau_{\rm gyro}$ via their mutual correlation with the
assumed stellar mass was not considered. With our Markov chain method we can
accurately account for the errors  in $\tau_{\rm iso}$ and $\tau_{\rm gyro}$
from the errors in T$_{\rm eff}$, $\rho_{\star}$ and [Fe/H] and calculate the
joint posterior distribution for these values (e.g., Fig.~\ref{CoRoT-13Fig}).
This enables us to accurately calculate a statistic like $p_{\tau}$ that has
greater statistical power than the method used by
\citeauthor{2014MNRAS.442.1844B}.

\citet{2009MNRAS.396.1789P} used the full sample of 41 transiting exoplanet
host stars known at that time to investigate whether there was empirical
evidence for tidal evolution in these planetary systems. He identified two
stars (HD~189733 and CoRoT-2) that show large excess rotation among the
late-type stars in this sample (T$_{\rm eff}<6000$\,K). This is consistent
with the values  of $p_{\tau} <0.1$ that we have derived for both these stars.
Additional constraints on the age of both these stars  are now available, so
we discuss these stars in more detail below. Pont's study also includes the
stars WASP-4 and WASP-5, but the only information on the rotation of these
stars available at them time was rotation velocities measured from spectral
line broadening comparable to the instrumental resolution. With the rotation
periods now available from photometry we find that both these stars show
some evidence for excess rotation ($p_{\tau} <0.1$). CoRoT-4 is also present
in both samples and shows no evidence for excess rotation in either study
($p_{\tau} =0.4$).

\subsection{Other age constraints}
 \citet{2014A&A...565L...1P} took a very different approach to testing whether
hot Jupiters affect the rotation rate of their host stars. They compared the
stellar coronal X-ray emission of 5 planet host stars with their companion
stars in wide binary systems and used this activity indicator to estimate the
age of both stars in each binary system. They found much higher magnetic
activity levels in HD~189733 and CoRoT-2 than in their 
companion stars.  The estimated age of HD~189733 based on the X-ray flux is
$\tau_{\rm X} =1$\,--\,2\,Gyr and for CoRoT-2 is $\tau_{\rm X} =
0.1$\,--\,0.3\,Gyr, whereas their companions both have $\tau_{\rm X} > 5$\,Gyr.
In contrast, there was no evidence for a difference in age between the
companion and the planet host star for $\tau$~Boo ($\tau_{\rm X} =
1$\,--\,2\,Gyr), $\nu$~And ($\tau_{\rm X} > 5$\,Gyr) or 55~Cnc ($\tau_{\rm X} >
5$\,Gyr). There is very good agreement between our estimate of $\tau_{\rm
gyro}$ and $\tau_{\rm X}$ for CoRoT-2 and $\tau_{\rm gyro}$ is consistent with
$\tau_{\rm X}$ in the case of 55~Cnc.  The probability that $\tau_{\rm iso}$
for CoRoT-2 calculated using our grid of standard stellar models  is
consistent with the value of $\tau_{\rm X}$ for its companion is $P(\tau_{\rm
iso} >5\,{\rm Gyr}) = 0.09$, i.e., if we assume that the value of $\tau_{\rm
X}$ for the companion is an accurate estimate of actual age of CoRoT-2 then the
isochronal age is likely to be an underestimate of the true age. Increasing the
helium abundance by about +0.04 or reducing $\alpha_{\rm MLT}$ by about 0.4,
or changing both these factors by about half as much, would be sufficient to
bring the $\tau_{\rm iso}$ and $\tau_{\rm X}$ into agreement.  For 55~Cnc,
both $\tau_{\rm iso}$ and $\tau_{\rm gyro}$ are consistent with each other and
with the lower limit $\tau_{\rm X} > 5$\,Gyr. \citeauthor{2014A&A...565L...1P}
conclude from their observations that the presence of hot Jupiters may inhibit
the spin-down of host stars with thick outer convective layers.  With a sample
of only 5 stars it is not yet possible to make precise estimates of what
fraction of planet host stars may be affected by their planetary companions or
what properties determine the strength of this interaction, but it is clear
that $\tau_{\rm gyro}$ is not a reliable estimate of the age for some planet
host stars.

 The lithium abundance at the surface of a star can provide useful constraints
on the age of a star, particularly for stars younger than about 600\,Myr where
there are good data in various open clusters that can be used to calibrate the
dependance of the lithium depletion rate with mass. We have compared the
observed lithium abundances for the stars listed in the Table~\ref{tgyroTable}
to the calibration data from \citet{2005A&A...442..615S} to see if the
resulting constraints on the age are consistent with the values of $\tau_{\rm
iso}$ and $\tau_{\rm gyro}$. In general we can only set a lower limit to the
age $\goa 0.6$\,Gyr that is consistent with both of these age estimates. A
notable exception is CoRoT-2, which has a surface lithium abundance that
implies an age for this star of $\approx 0.2$\,Gyr
\citep{2011A&A...532A...3S}. This agrees very well with the gyrochronological
age for this star. There does not appear to be any simple way to reconcile the
apparent young age of CoRoT-2 based on the available age indicators with the
lack of X-ray emission from its K-dwarf companion.

 Another case where there is evidence to support the gyrochronological age is
WASP-77. \cite{2013PASP..125...48M} used the projected rotational velocity of
the K-dwarf companion to WASP-77 to infer an gyrochronological age of
$0.4^{+0.3}_{-0.2}$\,Gyr, consistent with their estimate of
1.0$^{+0.5}_{-0.3}$\,Gyr for the gyrochronological age of WASP-77.
\cite{2013PASP..125...48M} used a different calibration for the
gyrochronological age to the one used here, but it is clear that the
gyrochronological age of companion to WASP-77 is inconsistent with the age for
WASP-77 estimated using a grid of standard stellar models. If we assume a
lower helium abundance for WASP-77 (Fig.~\ref{deltaFig}) we will obtain a
slightly lower estimate for the isochronal age, but it is not possible to
reconcile the gyrochronological age and isochronal age without using a helium
abundance less than the lower limit set by the primordial content of the
Universe. It is possible to reconcile the isochronal age with the
gyrochronological age by using a reduced value of the mixing-length parameter
$\alpha_{\rm MLT}$ much lower  than the solar-calibrated value. WASP-77 shows
chromospheric emission lines characteristic of magnetically active stars
despite its moderate rotation period, $P_{\rm rot} = 15.4$\,d. Reducing the
value of $\alpha_{\rm MLT}$  is consistent with the idea of magnetic
inhibition of convection that has been succesfully applied to explain the
radius anomaly in K-dwarfs, but extends the idea in this case to a G-type
star.\footnote{We take the dividing line between K-type and G-type dwarfs to
be 0.8$M_{\sun}$ in mass or T$_{\rm eff} = 5300$\,K.}

\subsection{Reliability of gyrochronological ages for single stars.}
 \citet{Meibom2015} have recently used {\it Kepler} photometry of 30 stars in the
open cluster NGC~6811 to measure their rotation periods. The calibration used
here \citep{2010ApJ...722..222B} predicts rotation periods for stars in good
agreement with the observed values at the age of the cluster (2.5\,Gyr) across
the entire mass range studied (0.85\,--\,1.25\,$M_{\sun}$).
\citet{2014ApJ...780..159E} criticised the method used by
\citet{2010ApJ...722..222B} to derive their calibration. We used Fig.~7 and
Fig.~8 of their paper to estimate ``by-eye'' the gyrochronological ages of all
the stars in our sample and found that these generally agree well with the
values derived here, particularly if the calibration using the Kawaler style
wind mass loss rate is used. For rapidly rotating stars the calibration of
\citet{2014ApJ...780..159E} provides only an upper limit of 0.5\,Gyr. This is
consistent with the young ages that we derive for these stars, but it may be
that these very young gyrochronological ages are not reliable to the precision
quoted here. This is unlikely to affect our conclusions substantially because
an upper limit of 0.5\,Gyr is generally sufficient to show that these rapidly
rotating stars have gyrochronological ages inconsistent with their isochronal
ages. The calibration of \citet{2014ApJ...780..159E} was also applied to the
data for stars in NGC~6811 by \citet{Meibom2015} and found to give rotation
periods that are too low by about 10\% for stars with masses $\approx
0.85\,M_{\sun}$ but that agree well with the observed rotation periods for
stars with masses near $1\,M_{\sun}$.

\section{Conclusions}
 By using new data and improved analysis methods we have shown that there is
now good evidence that some exoplanet host stars rotate more rapidly than
expected. For our sample of 28 transiting exoplanet host stars,
about half the sample have gyrochronological ages  that are significantly less
than their isochronal ages. In a few such cases there are independent
constraints on the age of the star that are consistent with the isochronal
age, which suggests that tidal spin-up of the host star has occured in these
systems. However, in several cases it is not clear that tidal interactions
between the star and the planet are responsible for this discrepancy. For some
K-type stars this is a result of the well-known radius anomaly that may be due
to magnetic inhibition of convection. We find some evidence that this anomaly
may also affect some of the rapidly rotating and/or magnetically active G-type
stars in our sample, either from independent age constraints on the age of the
star (WASP-77) or because it may be implausible that the strength of the tidal
interaction with the planet is strong enough to spin-up the star (CoRoT-6,
Kepler-63 and, perhaps, WASP-41). 

 Some planet-host stars (HAT-P-11 in particular) appear to be much denser than
predicted by stellar models. These stars may be significantly enhanced in
helium. This makes it difficult to assess the reliability of isochronal mass
and age estimates for these stars. There is currently no simple explanation
for the inconsistency between the young age of CoRoT-2 implied by stellar
models, gyrochronology, its X-ray flux and its high lithium abundance with the
very old age inferred  for its K-type companion based on its lack of X-ray
flux. 

 Our improved analysis methods have enabled us to show that there is now clear
evidence that the gyrochronological ages of some transiting exoplanet host
stars are significantly less than their isochronal ages. However, a careful
consideration of all the available data on a case-by-case basis shows that it
is not always clear that this is good evidence for tidal spin-up of the host
star by the planet.

\begin{acknowledgements}
 JS acknowledges financial support from the Science and Technology Facilities
Council (STFC) in the form of an Advanced Fellowship. AMS is supported by the
MICINN ESP2013-41268-R grant and the Generalitat of Catalunya program
SGR-1458. PM is grateful to Prof. Rob Jeffries for discussions about the lack
of  X-ray flux from the companion to CoRoT-2. We thank the anonymous
referee for their careful reading of the manuscript and comments that have
helped to improve this paper.
\end{acknowledgements}

\bibliographystyle{aa} 
\bibliography{wasp}
\end{document}